\documentclass[conference]{IEEEtran}
\IEEEoverridecommandlockouts
\usepackage{cite}
\usepackage{amsmath,amssymb,amsfonts}
\usepackage{algorithmic}
\usepackage{graphicx}
\usepackage{textcomp}
\usepackage{xcolor}
\usepackage{ctable}
\def\BibTeX{{\rm B\kern-.05em{\sc i\kern-.025em b}\kern-.08em
    T\kern-.1667em\lower.7ex\hbox{E}\kern-.125emX}}

\usepackage{microtype}

\usepackage{ulem}
\makeatletter
\def\ifEmpty#1{\def\@temp{#1}\ifx\@temp\@empty}
\makeatother

\usepackage{xspace}

\usepackage{amsmath,amsfonts,amssymb}
\usepackage{url} 

\newcommand{\FG}[1]{Figure~\ref{#1}}

\newcommand{\shah}{{\textstyle \amalg{\kern-4.pt\amalg}}}

\begin{document}

\title{The Illusion of Animal Body Ownership and Its Potential for Virtual Reality Games}

\author{\IEEEauthorblockN{1\textsuperscript{st} Andrey Krekhov}
\IEEEauthorblockA{\textit{Center for Visual Data Analysis} \\ \textit{and Computer Graphics (COVIDAG)} \\
\textit{University of Duisburg-Essen}\\
Duisburg, Germany \\
andrey.krekhov@uni-due.de}
\and
\IEEEauthorblockN{2\textsuperscript{nd} Sebastian Cmentowski}
\IEEEauthorblockA{\textit{Center for Visual Data Analysis} \\ \textit{and Computer Graphics (COVIDAG)} \\
\textit{University of Duisburg-Essen}\\
Duisburg, Germany \\
sebastian.cmentowski@uni-due.de}
\and
\IEEEauthorblockN{3\textsuperscript{rd} Jens Kr\"uger}
\IEEEauthorblockA{\textit{Center for Visual Data Analysis} \\ \textit{and Computer Graphics (COVIDAG)} \\
\textit{University of Duisburg-Essen}\\
Duisburg, Germany \\
jens.krueger@uni-due.de}
}

\IEEEpubid{\begin{minipage}{\textwidth}\ \\[12pt]
978-1-7281-1884-0/19/\$31.00 \copyright 2019 IEEE \end{minipage}}

\maketitle

\begin{abstract}
Virtual reality offers the unique possibility to experience a virtual representation as our own body. In contrast to previous research that predominantly studied this phenomenon for humanoid avatars, our work focuses on virtual animals. In this paper, we discuss different body tracking approaches to control creatures such as spiders or bats and the respective virtual body ownership effects. Our empirical results demonstrate that virtual body ownership is also applicable for nonhumanoids and can even outperform human-like avatars in certain cases. An additional survey confirms the general interest of people in creating such experiences and allows us to initiate a broad discussion regarding the applicability of animal embodiment for educational and entertainment purposes.
\end{abstract}


\section{Introduction}

Due to the depth of immersion, VR setups often excel at creating a strong bond between users and their virtual representations, the so-called avatars. That bond can be strong enough such that we start perceiving the avatar model as our own body---a phenomenon also known as the illusion of virtual body ownership (IVBO)~\cite{slater2010first}. Previous research agrees that VR is an efficient setup to induce IVBO experiences~\cite{slater2009inducing,slater2010first,waltemate2018impact}. However, the investigated scenarios have been centered mostly around humanoid avatars. Our paper aims at generalizing the IVBO discussion by considering virtual animals as candidates for an embodiment experience.

To provide a starting ground for future research regarding animal embodiment in VR, our work addresses the following question: \textit{Is IVBO applicable to nonhumanoid avatars, and, if so, what potential does that phenomenon have for VR applications and games in particular?} 

The primary contribution of our paper is the dedicated research of animal body ownership. Although prior work examined aspects such as the inclusion of additional limbs, tails, or wings, the question whether and how well we can embody virtual creatures remained unanswered. Our work provides a strong evidence that animal avatars can keep up and even outperform humanoid representations regarding IVBO. In our evaluation ($N=26$), we included a diversified set of animals to account for upright/flying species (bat), four-legged mammals (tiger), and arthropods (spider). Our experiment shows that even spiders (cf. \FG{fig:modes}), despite having a skeleton that significantly differs from ours, offer a similar degree of IVBO compared to humanoid avatars. Apart from the general assessment of IVBO, our paper proposes and discusses practical approaches to implement animal avatar control\footnote{supplementary video showcasing avatar controls: http://bit.ly/vranimals-cog} by, e.g., half-body tracking for non-upright avatars to reduce fatigue from crouching. We believe that our findings pave the way to the construction of a zoological IVBO framework in the future.


Our additional contribution is a discussion about the potential of VR animals. We conducted an online survey ($N=37$) that underpins the general interest of people in experiencing virtual animals---be it in educational documentaries or as protagonists in VR games. This survey supports our claim that VR has a potential regarding animal embodiment, resulting in application possibilities for a number of HCI areas such as games research.

\section{Motivation}
\label{section:application}

Who do we want to be in a game? Sorcerer, rogue, or warrior--these are default roles most gamers would think of. Even when a game offers more exotic choices for our avatar, we usually still get a humanoid representation. Thus, playable, realistic animals remain a rarity in common digital games, not to mention VR titles with only few exceptions such as \textit{Eagle Flight}~\cite{Eagle}. In our opinion, incorporating animals as player avatars into VR has the potential to unveil a set of novel game mechanics and maybe even lead to a ``beastly'' VR game genre. Furthermore, utilizing the abilities of animals such as flying as a bird or crawling as a spider could be significantly more engaging in VR due to the increased presence compared to non-VR games. Apart from entertainment, we suggest that embodying animal avatars could help us to better understand the behavior of a certain creature, e.g., in an educational documentary, and also increase our involvement with environmental issues~\cite{ahn2016experiencing}.

To capture the perspective of our society regarding these outlined applications, we administered an online survey with three sections: VR animals in general, animals in VR documentaries, and animals in VR games. Each section consisted of five questions about participants' experiences in that category and their general interest to give such a scenario a try. The questions were either yes/no, or on a 7-point Likert Scale ranging from 0 (\textit{totally disagree}) to 6 (\textit{totally agree}).

Thirty-seven subjects (21 female), aged 19 to 43 ($M = 26.43, SD = 5.67$), participated in the survey. Overall, 26 of the participants had prior experiences with VR, and 17 of them had already seen an animal in VR. However, only six subjects reported that they had had the chance to control a virtual animal. Our results indicate that the overall interest to try a VR application where animals play an important role is rather high ($M = 5.14, SD = 1.00$), and that subjects would like to observe, interact, and embody VR creatures (all $M > 4.50$).

Participants were keen on trying a VR game with an animal avatar ($M = 4.81, SD = 1.45$). Playing in third-person perspective ($M = 3.65, SD = 1.86$) was preferred less than in first-person ($M = 4.38, SD = 1.66$). However, a paired-samples t-test shows that these differences are not significant. The majority (33) had never seen a VR documentary about animals, but they said would like to try it ($M = 4.70, SD = 1.45$) and even embody a creature in such a documentary ($M = 4.24, SD = 1.80$). Participants also mostly agreed that embodying an animal might help them to better understand the animal's behavior ($M = 4.57, SD = 1.59$) and to increase their empathy toward that creature ($M = 4.89, SD = 1.45$).

Finally, the survey provided a multiple choice question as an opportunity for the subjects to tell us which animals they would like to experience in VR. The top three creature types were flying animals (birds, bats, etc.) with 32 votes, followed by typical mammals (lions, tigers, cats, dogs, etc.) with 30 votes, and by sea animals (dolphins, sharks, whales, etc.) with 25 votes. Combined with the results from our main IVBO study, we suppose that flying creatures indeed have the largest potential to fascinate users as embodiment targets in VR.

\section{Related Work on IVBO}

Immersive setups are capable of inducing the illusion of virtual body ownership (IVBO), also referred to as body transfer illusion, agency, or embodiment. IVBO~\cite{lugrin2015anthropomorphism} is an adaption of the effect of body ownership (BO), a term coined by Botvinick et al.~\cite{botvinick1998rubber}. The authors conducted an experiment to induce the so-called rubber hand illusion, in which they hid the participant’s real arm and replaced it with an artificial rubber limb. Both arms were then simultaneously stroked by a brush, which produced the illusion of owning the artificial arm. This effect has gained great publicity and was further researched by Tsakiris et al.~\cite{tsakiris2005rubber}. These results eventually led to the first neurocognitive model regarding body ownership~\cite{tsakiris2010my}, which emphasized the interplay between external sensory stimuli and the internal model of our own body. Additional studies extended
these finding to other limbs and whole-body representations~\cite{ehrsson2007experimental,petkova2008if,lenggenhager2007video}.

\begin{figure}[t]
\centering
\includegraphics[width=1.0\columnwidth]{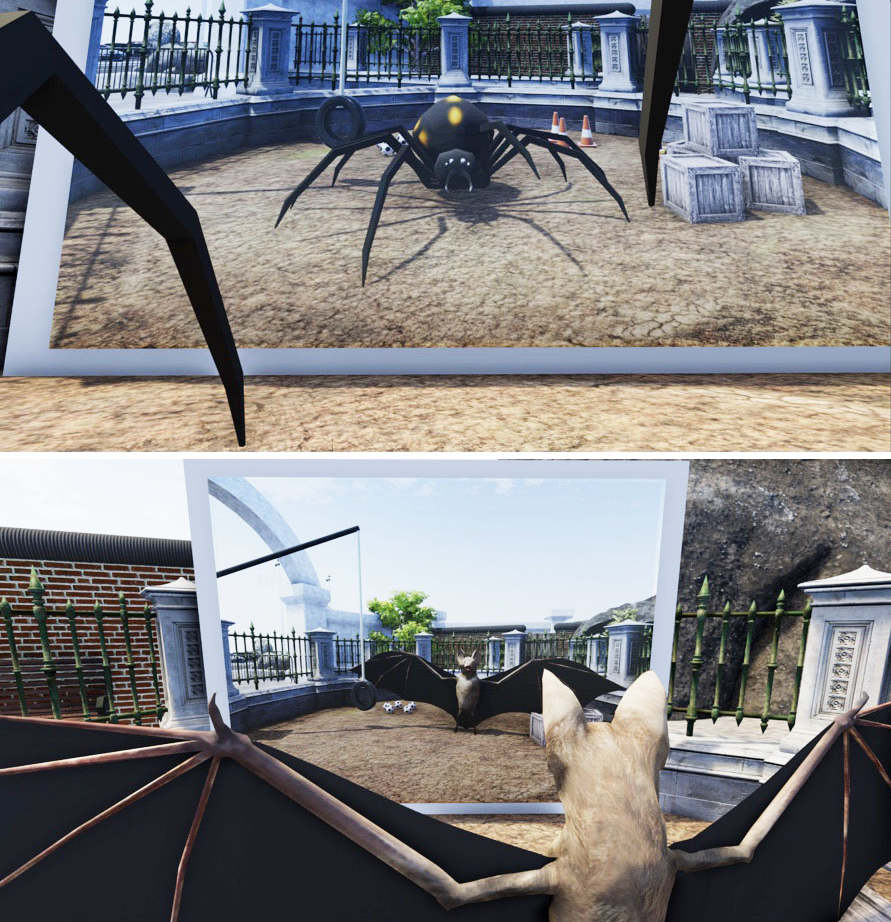}
\caption{Two of our avatars in first-person (top) and third-person (bottom) modes in front of a wall-sized mirror.}
\label{fig:modes}
\end{figure}

The effect of BO was initially transferred to virtual environments for arms by Slater et al.~\cite{slater2008towards} and entire bodies by Banakou et al.~\cite{banakou2013illusory}. However, these early studies used the original visuotactile stimulation introduced by Botvinick et al.~\cite{botvinick1998rubber}. Later research introduced sensorimotor cues, i.e., the tracking of hand and finger movement~\cite{sanchez2010virtual}, which was reported to be more important than visuotactile cues~\cite{slater2010first}. This finding is essential for VR setups as it releases possible experiments from the need for tactile stimulations. Furthermore, these two types of different cues are completed by the so-called visuoproprioceptive cues. These cues are a series of different body representations and include subdimensions such as perspective, body continuity, posture and alignment, appearance, and realism. These different subdimensions are listed in the correct order of influence on the effect of IVBO~\cite{slater2009inducing,slater2010first,perez2012my,maselli2013building}, and together are sufficient for inducing the illusion of body ownership~\cite{maselli2013building}. Moreover, Maselli et al.~\cite{maselli2013building} reported the necessity of a first-person perspective. In sum, IVBO is induced by correct visuoproprioceptive cues. Misalignments and visual errors can be compensated for through the weaker aspects of sensorimotor and visuotactile cues. However, this effect can be observed with anthropomorphic characters as well as realistic representations~\cite{lugrin2015anthropomorphism, lin2016need,jo2017impact}.

Riva et al.~\cite{riva2014interacting} illustrated the current interest in significantly altering the morphology of our virtual representation by the following question: \textit{But what if, instead of simply extending our morphology, a person could become something else- a bat perhaps or an animal so far removed from the human that it does not even have the same kind of skeleton— an invertebrate, like a lobster?} Especially for animals that have few characteristics in common with our human body, the approach of sensory substitution~\cite{bach2003sensory} is also a promising direction for IVBO research. For instance, we could replace the echolocation feature of a bat by visual or even tactile feedback in VR.

Recently, researchers have studied adapting and augmenting human bodies in VR. Kilteni et al.~\cite{kilteni2012extending} stretched the virtual arm up to four times its original length and were still able to confirm IVBO. These findings are in line with the work of Blom et al.~\cite{blom2014effects}, who reported that a strong spatial coincidence of real and virtual body part is not necessary for the illusion. Furthermore, researchers have determined that additional body parts are not necessarily destroying IVBO. Instead, it is possible to add a third arm and induce a double-touch feeling~\cite{ehrsson2009many,guterstam2011illusion}.

Apart from additional arms, other body parts have also been added successfully: Steptoe et al.~\cite{steptoe2013human} reported effects of IVBO upon attaching a virtual tail-like body extension to the user’s virtual character. The authors further discovered higher degrees of body ownership when synchronizing the tail movement with the real body. Another prominent example of body modification that could be relevant for embodying flying animals is virtual wings. In that area, Egeberg et al.~\cite{Egeberg:2016:EHB:2927929.2927940} proposed several ways to couple wing control with sensory feedback. Won et al.~\cite{won2015homuncular} further analyzed our ability to inhabit nonhumanoid avatars with additional body parts. Regarding realistic avatars, Waltemate et al.~\cite{waltemate2018impact} showed that customizable representations lead to significantly higher IVBO effects.

Strong effects of body ownership can produce multiple changes in the feeling or behavior of the user~\cite{jun2018full}, resembling the Proteus Effect by Yee et al.~\cite{yee2007proteus}. For instance, Peck et al.~\cite{peck2013putting} reported a significant reduction in racial bias when playing a black character. Additionally, virtual race can also affect the drumming style when playing virtual drums~\cite{kilteni2013drumming}. Other reactions are more childish feelings arising from child bodies~\cite{banakou2013illusory} or greater perceived stability due to a robotic self~\cite{lugrin2016avatar}. These findings demonstrate that IVBO is not just a one-way street but can be used to evoke specific feelings and attributes and possibly also change self-perception.

\begin{table*}
\renewcommand{\arraystretch}{1.3}
  \caption{Evaluated control modes for virtual animals.}
  \label{tab:controls}
  \begin{tabular}{l c p{10.5cm}}
    \toprule
    Mode & Evaluated avatars & Description\\
    \midrule
    first-person perspectives: & & \\
    \ \ \ \ \ \ full body \textbf{(FB)} & human, bat, spider, tiger & User's posture is mapped to the whole virtual body (cf. \FG{fig:mapping}. Mapping depends on the animal; see \FG{fig:mapping} for examples.\\[0.15cm]
    \ \ \ \ \ \ half body \textbf{(HB)} & spider, tiger & User's legs mapped to all limbs of an animal. \\[0.15cm]
    third-person perspectives: & & \\
    \ \ \ \ \ \ user centered \textbf{(3CAM)} & spider & The animated avatar is locked into a position in front of the user. \\[0.15cm]
    \ \ \ \ \ \ agent controlled \textbf{(3NAV)} & spider & The avatar is an autonomous agent following a target in front of the user. \\[0.15cm]
    \ \ \ \ \ \ avatar centered \textbf{(3FOL)}  & spider & The user is rotated around the animated avatar when turning. \\
    \bottomrule
  \end{tabular}
\end{table*}

We point readers to the recent work in progress by Roth et al.~\cite{roth2017alpha} regarding IVBO experience. In particular, the paper presented a IVBO questionnaire based on a fake mirror scenario study. The authors suggested acceptance, control, and change as the three factors that determine IVBO. In our experiments, we administered the proposed questionnaire as we were curious to see how it performs for animal avatars. Our research follows up on the works-in-progress paper by Krekhov et al.~\cite{krekhov2018anim}. The authors conducted a preliminary study with eight participants, and, by applying the alpha IVBO questionnaire~\cite{roth2017alpha}, concluded that IVBO might indeed work for animal avatars. We significantly extend that apparatus to gather more insights and to produce reliable results, and also to introduce additional surveys about virtual animals to explore the overall benefits of such research.

\section{Animal Embodiment}

As we can see from related work, body ownership requires as much sensory feedback as possible. Hence, if we want to evoke such experiences in a VR application, a simple gamepad control is probably not adequate.Prior research has shown that either proprioceptive cues or sensorimotor cues are necessary to induce proper levels of VBO. However, providing such cues is challenging for nonhuman characters as usually no straightforward control mapping exists between the participant and the virtual creature. 

In contrast to humans, animals come in various shapes, postures, and types, which makes it difficult to design a universal solution for avatar control. Therefore, our experiment includes multiple models combined with different types of control to gather diverse insights into animal embodiment.

Animal and human bodies differ in three main subdomains that are critical for successful body ownership: skeleton, posture, and shape, as can be seen in \FG{fig:mapping}. Certain animals, such as bats, share a human posture and skeleton but use scaled arms or legs and therefore vary in the natural shape, i.e., differ in terms of proportions. Other creatures such as tigers or dogs have an almost human skeleton, including the same number of limbs. However, they differ in the natural posture by walking on all fours. Finally, other species show a completely different skeleton and differ in the limb count. An appropriate example is a spider, which has eight legs attached to its head segment. To cover these different degrees of anthropomorphism, we have chosen tigers, spiders, and bats as our testbed species. In addition, we added a human avatar to compare our results with humanoid IVBO scores.

%

\subsection{Mapping Approaches}

We designed and evaluated multiple control modes and mapping approaches, as summarized in Table~\ref{tab:controls}. Even though prior work, e.g., by Debarba et al.~\cite{galvan2015characterizing}, underpins the superiority of first-person mappings regarding IVBO, that finding has not yet been confirmed for nonhuman embodiment. Hence, we decided to use both first-person and third-person perspectives (cf. \FG{fig:modes}) in our experiment to contribute to the perspective discussion.

The third-person perspective provides the advantage that subjects see their avatars standing right in front of them. However, that perspective is challenging when subjects rotate around themselves. For instance, in current non-VR games, the camera---or, in our case, the player---slides around the avatar to maintain the over-the-shoulder viewport.  This approach has been tested as one possible mode and named 3FOL. Another option is to use the subject as the rotational center and turn the animal around (3CAM). This mode is proposed to induce less cybersickness~\cite{laviola2000discussion} but lacks realism as the avatar slides sideways around the subject. Finally, this approach can be changed to enhance the visual quality by implementing a loose coupling: the animal avatar would be controlled by an agent trying to stay in front of the subject. This concept has the advantage that movement and rotation are chosen optimally to look natural while preserving the rotational center of 3CAM. We refer to this approach as 3NAV. In contrast to these three third-person perspectives, the first-person perspective is not affected by different rotational centers because the subject and the avatar share the same position.

Apart from different perspectives, our approaches also differ in the type of mapping that is applied. The tiger is usually walking on all fours. Hence, a subject imitating and becoming the animal could move the same with all four limbs being mapped to the tiger body. However, this full-body (FB) tracking is assumed to be somewhat exhausting as it forces participants to crouch on the floor. As an alternative, we introduce half-body (HB) tracking : the subjects stand or walk in an upright position, watch through the eyes of their animal, and have their lower body mapped to all of the animal’s limbs. For instance, in case of a tiger, one human leg corresponds to two of the animal's pawns. This variation preserves the amount of sensory feedback while reducing the necessary physical effort. Another approach---the one we used for the third-person perspectives---is to avoid posture tracking and replace it with predefined avatar animations, only keeping the subject's position and orientation in sync.


\begin{figure*}[t!]
\centering
\includegraphics[width=1.87\columnwidth]{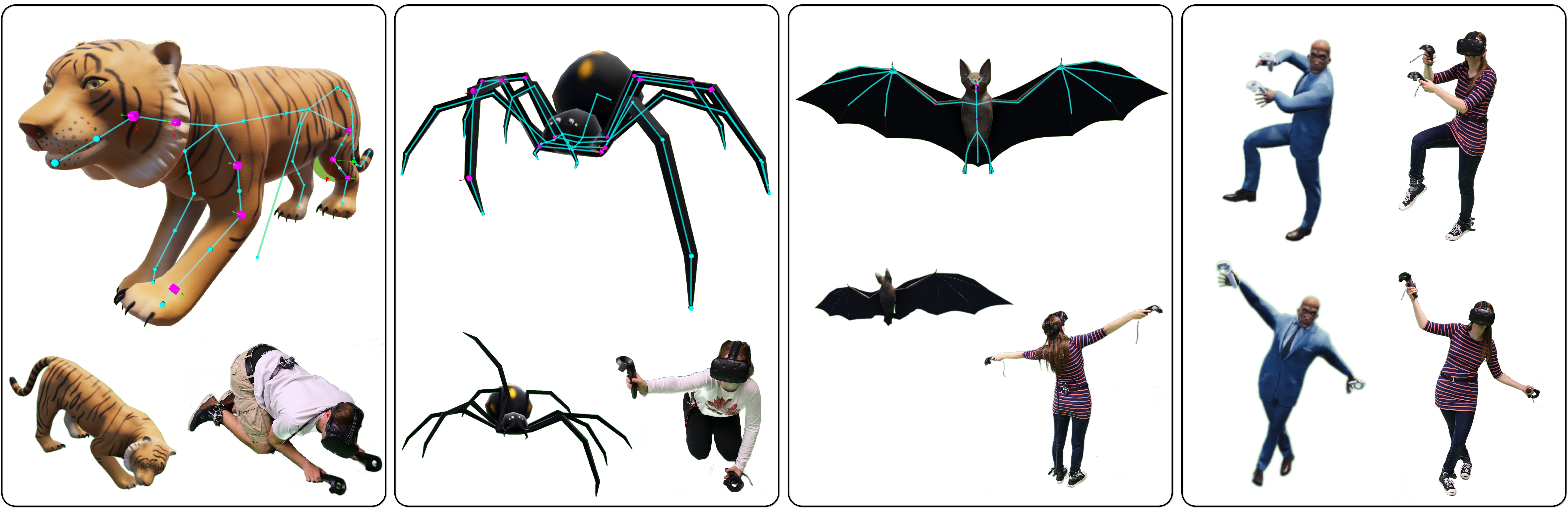}
\caption{Three virtual animals, their controls in FB mode, and the human avatar that was used as the reference for IVBO comparisons. The animals were chosen such that they differ from humanoids in IVBO-critical domains, i.e., shape (bat), skeleton (spider), and posture (tiger, spider).}
\label{fig:mapping}
\end{figure*}

\subsection{Testbed Scenario}

We utilized a combination of Unity3D and HTC Vive, including additional Vive trackers positioned at the hip and both ankles to enable full-body positional tracking. The HB and FB modes required custom avatar poses depending on tracker positions and rotations. Therefore, we experimented with different approaches based on inverse kinematics (IK)~\cite{buss2004introduction}. Physical models typically used for ragdoll systems and iterative solvers tended to jitter and flicker upon combining them with the VR tracking. As these issues were partially caused by unavoidable tracking errors, these approaches did not suit the situation. Instead, we applied a combination of closed-form and iterative solvers to achieve more stability at the cost of limited rotational movement. 

As depicted in \FG{fig:scene}, we placed our experiment in a stereotypical zoo where the participants were locked inside an arena-like cage filled with different interactive items such as cans, crates, or tires. Moreover, we installed a virtual wall-sized mirror to enhance the VBO illusion~\cite{latoschik2016fakemi}. So we relied on the same testbed scene for all conditions, animals were scaled to roughly equal, human-like dimensions.

\subsection{Hypotheses and Research Questions}

Our main goal is to explore animal embodiment by evaluating the five proposed mapping approaches with different animals. We want to see how potential users perceive the different control modes and what they like or dislike about our animal avatars in VR. Furthermore, we hypothesize that, similar to humanoid IVBO findings~\cite{galvan2015characterizing}, third-person modes for animals are inferior to the first-person perspective. To summarize, our questions and hypotheses are the following:
\begin{itemize}
  \setlength{\itemsep}{2pt}
  \setlength{\parskip}{0pt}
  \setlength{\parsep}{0pt}
\item RQ1: How do first-person modes (FB and HB) for animals perform regarding IVBO compared to a human avatar?
\item RQ2: Do our creature types differ regarding IVBO and user valuation?
\item RQ3: Is there any difference between FB and HB for the same animal?
\item H1: First-person modes significantly outperform third-person modes regarding induced IVBO.
\end{itemize}

\subsection{Procedure and Applied Measures}

We conducted a within-subjects study in our VR lab and tested the following conditions: FB human (as reference for IVBO), FB spider, FB tiger, FB bat, HB spider, HB tiger, 3CAM spider, 3NAV spider, and 3FOL spider. We excluded the HB bat case because that animal can be controlled in an upright pose in FB. Thus, we do not see any advantage to using the lower body only. We limited the third-person modes to one animal because these approaches behave the same for all animals.

Upon the participants' arrival, we administered a general questionnaire assessing age, gender, digital gaming behavior, and prior experiences with VR systems. For each condition, we told the participants to move around in the virtual arena and experiment with their virtual representation. For instance, subjects were able to move and drag various objects, such as crates, pylons, and tires. Subjects stayed in the virtual world for around five minutes for each condition. This duration is a typical choice for IVBO studies~\cite{tsakiris2005rubber} despite the finding that even 15 seconds may be enough to induce body ownership~\cite{lloyd2007spatial}. 

We decided against performing threat tests for capturing IVBO, as the sequence effects in our case would be too significant. Note there is no unified procedure for measuring IVBO and a threat test is not the only possibility~\cite{kilteni2012sense,roth2017alpha}. Instead, we decided to use the alpha IVBO questionnaire by Roth et al.~\cite{roth2017alpha}, and also checked its reliability by calculating Cronbach’s alpha for all subscales (all alphas $> 0.81$).

We administered the alpha IVBO questionnaire after each condition. Answers were captured on a 7-point Likert Scale ranging from 0 (\textit{totally disagree}) to 6 (\textit{totally agree}). In particular, the questionnaire captures the three dimensions acceptance, control, and change. Acceptance reflects self-attribution and owning of the virtual body by statements such as: \textit{I felt as if the body parts I looked upon were my body parts}. Control mostly focuses on the correct feedback and agency. One example is: \textit{I felt as if I was causing the movement I saw in the virtual mirror}. Finally, change measures self-perception and is usually triggered when the avatar differs much from the user. Three subitems focus on changes during the experiment (e.g., \textit{At a time during the experiment I felt as if my real body changed in its shape, and/or texture}), whereas another three subitems capture after-effects (e.g., \textit{I felt an after-effect as if my body had become taller/smaller}).

%
%
%
%

We extended the questionnaire by additional custom questions and statements to capture fascination (\textit{The overall experience was fascinating}), ease of control (\textit{I coped with the control of the avatar}), and fatigue (\textit{Controlling the avatar was exhausting}). We used the same scales as for the IVBO questions. Furthermore, we conducted semistructured interviews after each control mode, i.e., after FB, HB, 3CAM, 3NAV, and 3FOL. In particular, we asked subjects what they liked best/least and why, whether they could imagine such controls in a VR game, and how we could further enhance that mode. Upon completion of all conditions, participants had the chance to provide general feedback regarding animal avatars and tell us their favorite animals to be included in the next experiments.

Twenty-six subjects (13 female) with a mean age of $23.46$ ($SD=7.06$) participated in our study. Most participants (21) reported playing digital games at least a few times a month, and the majority (21) had used VR gaming systems. 

To address our hypothesis and research questions, we compare all nine conditions regarding their IVBO performance for acceptance, control, and change. All investigated parameters were approximately normally distributed according to Kolmogorov-Smirnov tests. Hence, we used one-way repeated measures ANOVA to compare the measured IVBO  values outlined in \FG{fig:diagram}. The outcomes differed significantly in all three dimensions, i.e., acceptance, \textit{F}~(4.86, 122.14)~=~18.23, \textit{p}~<~.001, control, \textit{F}~(3.64, 90.95)~=~18.54, \textit{p}~<~.001, and change, \textit{F}~(4.11, 102.73)~=~14.54, \textit{p}~<~.001. Post hoc Bonferroni tests provided additional insights into these differences:

\textbf{Acceptance:} the human avatar ($M = 2.79, SD = 1.31$) was rated significantly lower than FB bat ($M = 4.33, SD = 1.10$) and HB spider ($M = 3.63, SD = 1.29$) with $p < .01$ in both cases. 

\textbf{Control:} FB bat ($M = 5.11, SD = 0.82$) achieved significantly higher scores than all other modes, whereas 3NAV ($M = 2.41, SD = 1.38$) and 3FOL ($M = 2.55, SD = 1.22$) performed significantly worse than each first-person mode (all $p < .05$). 

\begin{figure}[b]
\centering
\includegraphics[width=1.0\columnwidth]{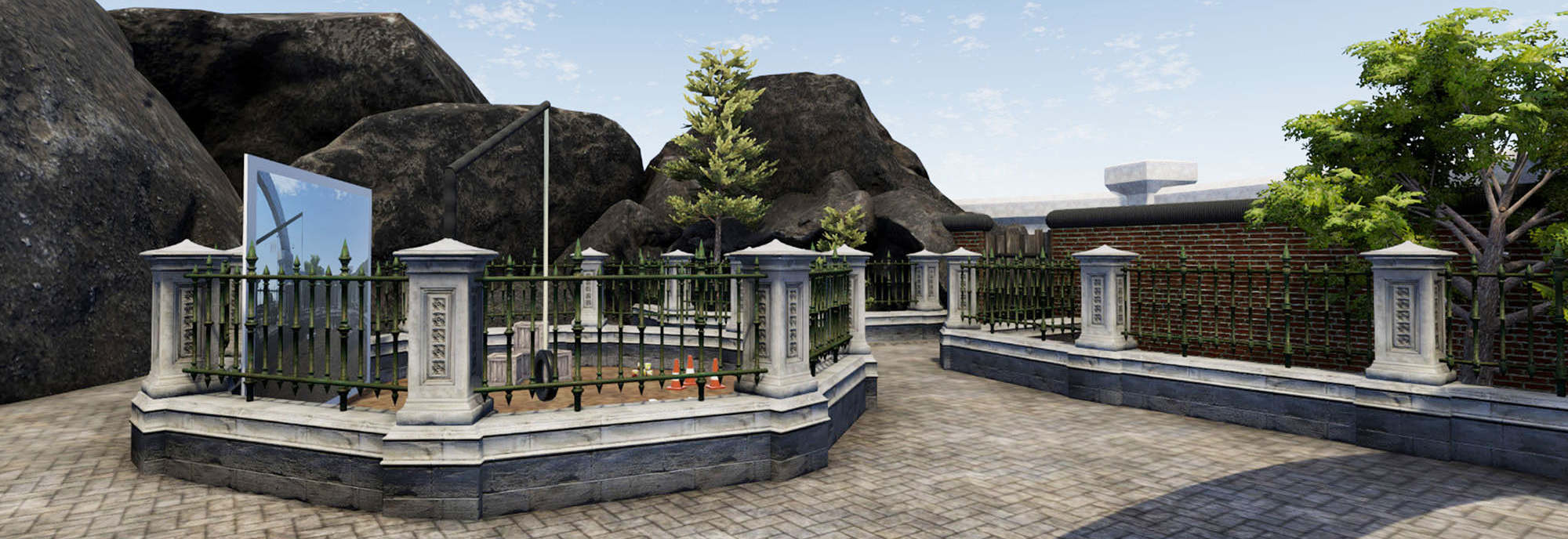}
\caption{We chose a virtual zoo for our testbed scenario. The cage is equipped with a wall-sized mirror to enhance IVBO.}
\label{fig:scene}
\end{figure}

\subsection{Results}

\begin{figure*}[t!]
\centering
\includegraphics[width=2\columnwidth]{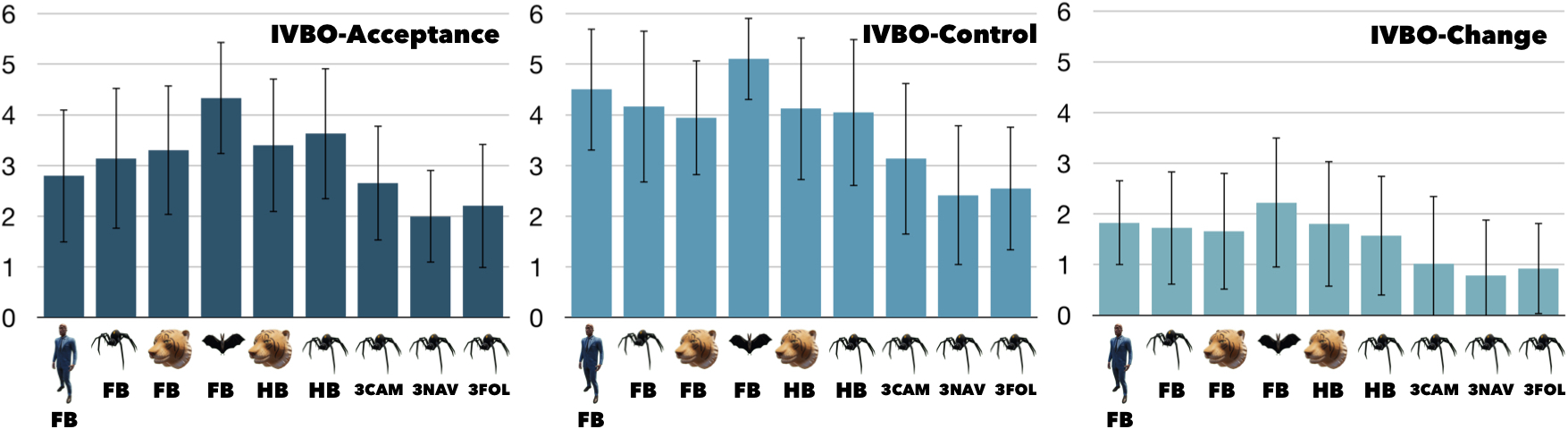}
\caption{Mean scores and standard deviations for the three IVBO dimensions: acceptance, control, and change.}
\label{fig:diagram}
\end{figure*}

\textbf{Change:} All modes had rather low values, as can be seen in \FG{fig:diagram}. FB bat ($M = 2.22, SD = 1.28$) was rated significantly better (all $p < .01$) than HB tiger ($M = 1.57, SD = 1.18$), 3CAM ($M = 1.02, SD = 1.34$), 3NAV ($M = 0.79, SD = 1.10$), and 3FOL ($M = 0.92, SD = 0.90$). Also, all first-person modes significantly outperformed 3NAV and 3FOL (all $p < .05$).

To create a better picture for RQ2, we also considered our custom questions and statements summarized in \FG{fig:diagram2}. Our conditions significantly differed regarding fascination, \textit{F}~(3.68, 92.08)~=~2.99, \textit{p}~=~.026, ease of control, \textit{F}~(5.10, 127.58)~=~8.58, \textit{p}~<~.001, and fatigue, \textit{F}~(5.43, 135.76)~=~13.46, \textit{p}~<~.001.  Post hoc Bonferroni tests revealed the following details:

\textbf{Fascination:} FB bat ($M = 5.27, SD = 0.78$) significantly outperformed (all $p < .05$) FB human ($M = 4.42, SD = 1.24$), HB tiger ($M = 4.08, SD = 1.55$), 3CAM ($M = 4.27, SD = 1.51$) and 3FOL ($M = 3.88, SD = 1.80$).

\textbf{Ease of Control:} Again, FB bat ($M = 5.38, SD = 0.75$) was perceived very positively and had significantly higher scores (all $p < .01$) than all other modes except 3CAM ($M = 4.65, SD = 1.47$), which was ranked second. In contrast, 3FOL ($M = 3.04, SD = 1.71$) produced most control difficulties, which resulted in significantly lower scores (all $p < .05$) than FB human, FB bat, HB spider, and 3CAM. 

\textbf{Fatigue:} Similarly, 3FOL ($M = 3.65, SD = 2.00$) was most exhausting and performed significantly worse (all $p < .05$) than all modes except FB tiger ($M = 3.58, SD = 1.39$) and FB spider ($M = 3.35, SD = 1.55$). The two latter modes were also rated significantly inferior (all $p < .05$) to the remaining modes, which all stayed below 2 as mean value.

\begin{figure*}[t!]
\centering
\includegraphics[width=2\columnwidth]{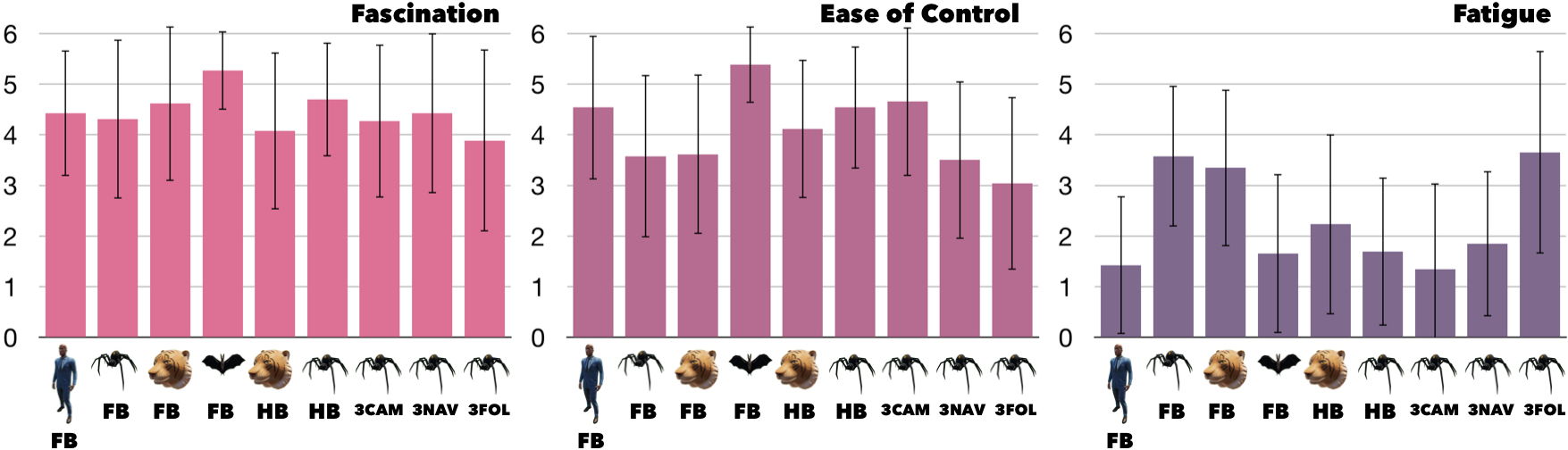}
\caption{Mean scores and standard deviations for fascination (\textit{The overall experience was fascinating}), ease of control (\textit{I coped with the control of the avatar}), and fatigue (\textit{Controlling the avatar was exhausting}).}
\label{fig:diagram2}
\end{figure*}


\subsection{Discussion}



\subsubsection{How do first-person modes (FB and HB) for animals perform regarding IVBO compared to a human avatar? - \textbf{RQ1}}
In all three IVBO dimensions, ANOVA did not reveal any significant advantages of FB human over the animal first-person modes. On the contrary, for acceptance and control, the humanoid representation was significantly outperformed by FB bat, and, for acceptance, also by HB spider. Hence, our main observation is that IVBO should be applicable for nonhumanoid avatars that differ in shape, skeleton, or posture from our human body. 

However, we are aware that the appearance of the human avatar has also a strong impact on IVBO. For instance, customizing that representation~\cite{waltemate2018impact} could produce significantly higher IVBO scores for that condition. Thus, we do not want to exaggerate the generality of our finding, and rather state that animal IVBO has the potential to keep up with humanoid IVBO and, thus, is worth further, more detailed investigations.

\subsubsection{Do our creature types differ regarding IVBO and user valuation? - \textbf{RQ2}}

In our case, the clear ``winner'' regarding IVBO scores and our custom questions is the bat, a creature type that mostly differs in shape but maintains similar posture and skeleton compared to our body. This finding might be an indication that animals with human-like, upright postures are more suited for IVBO effects. The quantitative results also align with the interview feedback that we got during the evaluation: \textit{``The bat behaved exactly how I expected and it was intriguing to precisely control my wing movements because it appeared realistic to me''(\textbf{P7})}. Subjects often expressed their desire to utilize the flying capability: \textit{``I could feel more like a giant bat if I could fly by moving my arms and maybe lean forward to accelerate''(\textbf{P4})}.

Another finding is the different perceptions of FB spider and FB tiger modes. Participants reported that the tiger felt less engaging, and we recorded several similar statements such as the following: \textit{``The forepaws were too short, they even felt shorter than my real arms and I could not do much with them''(\textbf{P2})}. Of course, tiger paws are not shorter, but the tiger head position leads to a distorted perspective. Hence, we suppose that perceiving virtual limbs as shorter than our real limbs feels rather limiting, whereas longer virtual limbs are classified as useful tools that enhance our interaction space. This finding would also explain the supremacy of FB bat mode with wings as extended tools: \textit{``The long arms of the bat felt a bit like two long sticks I could use to reach more items''(\textbf{P18})}.

\subsubsection{Is there any difference between FB and HB for the same animal? - \textbf{RQ3}} Our experiment did not reveal any significant differences in that regard, which is surprising because HB reflects only half of our posture changes. HB spider overall achieved positive ratings that were close to FB bat and even significantly outperformed FB human regarding acceptance.

\begin{figure}[b]
\centering
\includegraphics[width=1.0\columnwidth]{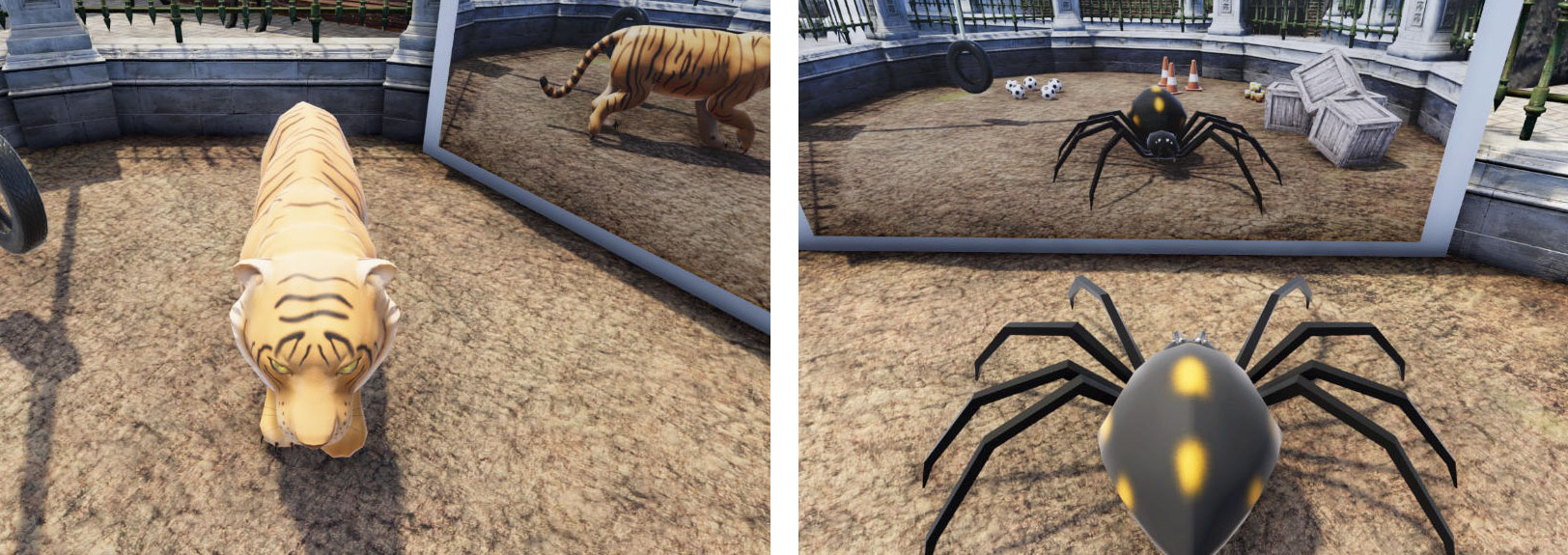}
\caption{A difference of 3NAV (left) to 3CAM and 3FOL (right) occurs when subjects walk backwards: the avatar is suddenly facing and ``chasing'' them.
}
\label{fig:chase}
\end{figure}
Our custom question related to fatigue revealed that subjects perceived FB modes for spider and tiger control to be significantly more exhausting than their HB counterparts because they had to kneel and crouch on a yoga mat. For such types of animals, HB modes seem to be more promising because they expose the same amount of IVBO without being aggravating. However, one disadvantage of HB is the less direct mapping, i.e., subjects \textit{``felt limited regarding possible interactions as it is difficult to forecast the avatar behavior sometimes''(\textbf{P2})}. Regarding missing control in HB, two participants mentioned that, in case of a spider, they would like \textit{``to control each limb separately, maybe even with finger movements''(\textbf{P5})}.

\subsubsection{First-person modes (FB and HB) significantly outperform third-person modes (3CAM, 3NAV, 3FOL) regarding induced IVBO - \textbf{H1}} Overall, all third-person modes achieved rather low scores in all IVBO dimensions. In particular, for control and change, the 3NAV and 3FOL modes were significantly outperformed by all first-person perspectives, which mostly supports our hypothesis and is in line with humanoid research~\cite{galvan2015characterizing, maselli2013building}. Hence, if a higher IVBO is desired, controlling an animal in first-person mode is advantageous.

The scores for 3CAM were not significantly lower compared to first-person modes, which renders that approach a viable alternative if first-person is not possible. In the interview, most subjects reported slightly preferring the 3CAM mode over 3NAV and 3FOL. In the 3NAV condition, subjects perceived the avatar to be \textit{``controlled telepathically''(\textbf{P8})} and to \textit{``orbit the player''(\textbf{P2})}. One participant was surprised at one point (cf. \FG{fig:chase}): \textit{``When I walked backward, the spider suddenly looked at me and seemed to chase me''(\textbf{P1})}. We suggest that 3CAM and 3NAV are both suited for VR applications, yet 3NAV resembles more companion-like behavior rather than an avatar representation. We do not recommend using the 3FOL mode, as it is capable of evoking dizziness, as was confirmed by two participants. Especially regarding the question about how exhausting the control was, 3FOL performed significantly worse compared to other third-person perspectives. Hence, even though it is widely used in non-VR games, we do not see any notable advantages of 3FOL in a VR setup.

\subsubsection{Design Implications} The outcomes of our experiment allow the formulation of design considerations for further research. In first place, we argue that a 1:1 full-body mapping is not a key requirement for IVBO, as half-body approaches often achieved similarly high scores. This observation is especially important for the design of animals with significantly different skeletons that cannot be mapped to our human anthropology, as we can still induce the IVBO effect under such conditions. 

In general, we note that half-body approaches that map one of our legs to multiple animal limbs should be considered instead of forcing users in a non-upright position such as crouching. Half-body maximizes the IVBO effect compared to less direct mapping modes, yet removes the discomfort induced by full-body controls. However, using our legs only limits the interaction precision, and we recommend mapping fine-grained manipulation tasks to our hands. For instance, we could map the two front limbs of the spider to our arms, which allows users to execute precise actions such as holding objects.

Another observation is related to the choice of perspective. Although third-person approaches were inferior to first-person modes regarding IVBO, we argue that third-person offers a viable option for rapid prototyping of VR animal applications. First-person modes presume a precise motion mapping to perform well, which usually requires tuned IK solutions and knowledge of animal kinesiology. In contrast,  third-person controls---probably due to the weaker IVBO---can rely on simple, predefined avatar animations: in our experiment, participants have not noticed any difference nor reported any resentment related to the unsynchronized movements.

\section{Conclusion and Future Work}

Backed by our supplementary studies, we underpinned the large potential of animal avatars for VR research and applications, be it for education or entertainment. To provide a starting point for future research, we proposed a number of different control modes for upright/flying species, four-legged mammals, and arthropods. Our evaluation revealed that IVBO can be considered for nonhumanoid avatars and led us to a first set of design implications in that area. 

We conclude that half-body tracking is a viable alternative to control animals that are not in an upright position as it offers a promising trade-off between fatigue and IVBO. For that reason, we suggest examining such half-body approaches in more depth. To provide higher degrees of control, a combination with sensory substitution~\cite{bach2003sensory} might be a viable approach for future research. Finally, as desired by the majority of participants, we propose to enhance the avatars with appropriate capabilities such as flying and see how this would impact IVBO. Hereby, the ultimate goal is the construction of a zoological IVBO framework that would support researchers and practitioners in designing meaningful virtual animals.


\bibliographystyle{IEEEtran}
\bibliography{IEEEabrv,vranim-bibliography}

\end{document}